\documentclass{article}
\usepackage[a4paper, hmargin=1.5cm, vmargin={2cm,2cm}]{geometry}
\usepackage[utf8]{inputenc}
\usepackage{amsmath,amssymb}
\usepackage{xcolor}

\usepackage[fontsize=14pt]{scrextend}
\usepackage{graphicx}
\usepackage{hyperref}
\usepackage[T1]{fontenc}
\usepackage[font=small,labelfont=bf,tableposition=top]{caption}
\usepackage{pythonhighlight}
\usepackage{amsthm}
\usepackage{indentfirst}
\usepackage{mathtools}

\title{The effect of normal electric fields on the Stokes drift}
\author{ Luiz P. Palacio$^{1}$, Marcelo V. Flamarion$^{2}$,  Tao Gao$^{3}$,  Roberto Ribeiro-Jr$^{1}$}
\date{}

\begin{document}
	\maketitle
	\begin{center}
		
			{\footnotesize $^1$UFPR/Federal University of Paran\'a,  Departamento de Matem\'atica, Centro Polit\'ecnico, Jardim das Am\'ericas, Caixa Postal 19081, Curitiba, PR, 81531-980, Brazil  \\
			luizpalacio@ufpr.br/ robertoribeiro@ufpr.br } 
		
			{\footnotesize $^2$Departamento Ciencias-Secci{\' o}n Matem{\' a}ticas, Pontificia Universidad Cat{\' o}lica del Per{\' u}, Av. Universitaria 1801, San Miguel 15088, Lima, Peru \\
			
mvellosoflamarionvasconcellos@pucp.edu.pe} \\

{\footnotesize $^3${\color{black}School of Mathematics, Statistics and Actuarial Science}, University of Essex, Colchester CO4 3SQ, United Kingdom \\
			
t.gao@essex.ac.uk} \\

	\end{center}
	
	\begin{abstract}
		
\noindent In periodic wave motion, particles beneath the wave undergo a drift in the direction of wave propagation, a phenomenon known as Stokes drift. While extensive research has been conducted on Stokes drift in water wave flows, its counterpart in electrohydrodynamic flows remains relatively unexplored. Addressing this gap, we investigate Stokes drift beneath periodic traveling irrotational waves on a dielectric fluid under the effect of normal electric fields. Through numerical simulations utilizing conformal mapping, we compute particle trajectories and analyze the resultant Stokes drift behaviors beneath periodic traveling waves. Our findings indicate that variations in the electric field impact particle velocities while maintaining trajectory shapes. Moreover, the kinetic energy associated with a particle depends on its depth location and is a non-decreasing convex function in a laboratory frame and a constant in a moving frame,  as observed in water wave flows.
		
	\end{abstract}
	
\section{Introduction}

The study of particle trajectories beneath a surface water wave dates back to 1839, and it is a problem of great physical and mathematical interest, being relevant for various applications, such as submarine operations, the drift of particles including oil spills, gas bubbles, suspended sediment, and biological materials \cite{Mostafa:2015, NachbinRibeiro:2014}.

Historically, George Green \cite{Green:1839}  pioneered the study of particle trajectories beneath a free surface wave. He proved that the particle trajectories beneath a linear wave in the deep-water regime are circular and closed, with the radius decreasing exponentially with depth. Shortly after, George Biddell Airy \cite{Airy:1841} deduced that in finite depth, these paths are elliptical. Green's work inspired one of the most important figures in fluid dynamics, George Gabriel Stokes, who was mentored by William Hopkins at the University of Cambridge. In autobiographical notes, Stokes recalled \cite{Craik:2004}:

\begin{quote}
	
	I thought I would try my hand at original research; and, following a suggestion
	made to me by Mr. Hopkins while reading for my degree, I took up the subject
	of Hydrodynamics, then at rather a low ebb in the general reading of the place,
	notwithstanding that George Green, who had done admirable work in this and
	other departments, was resident in the University till he died. (Larmor 1907, p.8)
\end{quote}

Stokes \cite{Stokes:1847} deepened Green's studies using successive approximations, showing that in both shallow and deep water regimes, the particles beneath a wave undergo a drift in the direction of wave propagation, which was later called \textit{Stokes Drift}. In this way, he conjectured that particle paths are not closed.

Years later, additional findings related to Stokes' conjecture were made. Ursell \cite{Ursell:1953} proved that the paths of particles are not closed for waves in either shallow or deep water regimes with uniform depth. Longuet-Higgins \cite{Longuet-Higgins:1986} conducted an experimental study to determine the trajectory of a particle on the surface and found loop-shaped orbits with a drift in the direction of wave propagation. Constantin \cite{Constantin:2006} provided a rigorous proof for Stokes' conjecture; in other words, he proved that particles beneath a wave undergo a drift in the direction of wave propagation in both shallow and deep water regimes and showed the geometry of the particle paths, which are loop-shaped. Constantin and Villari \cite{ConstantinVillari:2008} analyzed a more specific case involving linear waves, where the velocity field of the fluid is known, and obtained the same result as in \cite{Constantin:2006}. Constantin and Strauss \cite{ConstantinStrauss:2010} studied particle trajectories originating from the interaction of the free surface wave with a uniform current, and in this case, they proved that it is possible for the particles not to undergo a drift.

In more recent studies, significant progress has been made on various problems involving particle trajectories beneath free surface waves in reduced models \cite{Kalisch:2013, Kalisch:2012, Flamarion:2023, Flamarion:2023-2, Khorsand:2014}. Some authors have focused on particle trajectories beneath irrotational Stokes waves. Carter et al. \cite{KalischCarterCurtis:2019} primarily used the Nonlinear Schr{\"o}dinger equations to describe the surface of a Stokes wave and study the paths of the particles beneath this wave, obtaining similar results to those for a Stokes wave described by the Euler equations. Vanneste and Young \cite{VannesteYoung:2022} show that the Stokes drift can be decomposed into a solenoidal component, which simplifies the analysis of the Stokes drift and particle paths beneath Stokes waves. Other authors consider waves with constant vorticity, representing a realistic flow when waves are long compared to the depth or short compared to the length scale of the vorticity distribution \cite{TelesDaSilvaPeregrine:1988}. Van den Bremer and Breivik \cite{BremerBreivik:2018} studied Stokes drift through experimental studies using photographic techniques in the laboratory to compare theoretical and experimental results. They also discussed three main areas of application for Stokes drift: in the coastal zone, in Eulerian models, and in models of tracer transport, such as oil and plastic pollution. Abrashkin and Pelinovsky \cite{AbrashkinPelinovsky:2018} analyzed the Stokes drift beneath two types of waves, Stokes waves and Gerstner waves, and showed a relationship between Gerstner waves, Stokes waves, and Stokes drift. Specifically, the quadratic approximation of particle trajectories beneath a Stokes wave is a superposition of the vorticity flow of the Gerstner wave and the shear flow of the Stokes drift. Weber \cite{Weber:2019} demonstrated that the Stokes drift beneath a Gerstner wave is zero using a nonlinear Lagrangian formulation.

All the aforementioned studies focus on the behaviour of particles beneath water waves. A relevant field, called electrohydrodynamics (EHD), studies the coupling of charged fluid motion with electric fields, where an interface between two fluids is usually concerned in practical scenarios. It has numerous applications in chemical engineering, e.g. coating processes \cite{KS:1997, GriffingBankoffMiksisSchluter:2006}, cooling systems in high-power devices \cite{GhoshalMiner:2003}, the electrospray technology \cite{TD:1969} etc. The readers may refer to \cite{CCY:2003,P:2019} for more details. There were two major early achievements in the literature on interfacial waves by Taylor \& McEwan \cite{TM:1965} and Melcher \& Schwarz \cite{MS:1968} respectively. The former work theoretically and experimentally demonstrated that the interface between a conducting fluid and a dielectric could be destabilized due to normal electric fields perpendicular to the undisturbed interface (i.e. vertically). Meanwhile, the latter work considered the problem with tangential electric fields parallel to the undisturbed surface and showed by performing a linear stability analysis that short waves could be regularised under this setting. Since then, many authors have continued to investigate the role of electric fields in (de)stabilizing various interfacial fluid configurations. For the problem of vertical electric fields, the most general setting concerns two immiscible fluids, with an interface in between, of perfect dielectric of different electric permittivities (see \cite{D:2020} for details). Some assumptions are usually made to simplify the problem. One common setup is to consider the upper layer to be a hydrodynamically passive region of dielectric and the lower layer to be a conducting fluid, see e.g. \cite{DC:1974, DC:1988, DC:2006, DC:2007, DC:2013, DC:2015, DC:2017, DC:2022}. The other common assumption is to let the upper layer be conducting gas while the lower layer is a dielectric fluid, which reduces the physical configuration to a one-layered problem, see e.g. \cite{CD:2018,CD:2019,FlamarionGaoRibeiroDoak:2022,FlamarionGaoRibeiro:2023}. A direct question in the present context is how the electric fields affect the Stokes drift. To fill this gap in the literature, we investigate the Stokes drift of a charged particle beneath an irrotational Stokes wave under normal electric fields in the same physical configuration as in \cite{CD:2019} in this paper.  More specifically, we use conformal mapping and pseudo-spectral numerical methods to compute the particle trajectories and evaluate the Stokes drift.


The article is structured as follows: The mathematical formulation is presented in Section \ref{section2}. The linear case and the properties of Stokes drift are reviewed in Section \ref{sec-linear}. The numerical method for the nonlinear problem is presented in Section \ref{sec-nonlinear}. Results are presented in Section \ref{sec-results}. Finally, concluding remarks are provided in Section \ref{sec-conc}.

\section{Mathematical Formulation}\label{section2}

We consider an inviscid and incompressible dielectric fluid of permittivity $\epsilon_0$ bounded by wall electrodes on top and bottom, {\color{black}{imposed with a constant voltage difference inbetween}}, and surrounded by a conducting gas. In this case, the fluid is conditioned to the action of a normal electric field ($\Vec{E} = \nabla V$), {\color{black}{in which $V$ is a voltage potential. We may let $V=0$ on top and $V=-V_0$ on bottom without losing generality.}} 
\begin{figure}[h!]
	\centering
	\includegraphics[scale = 0.5]{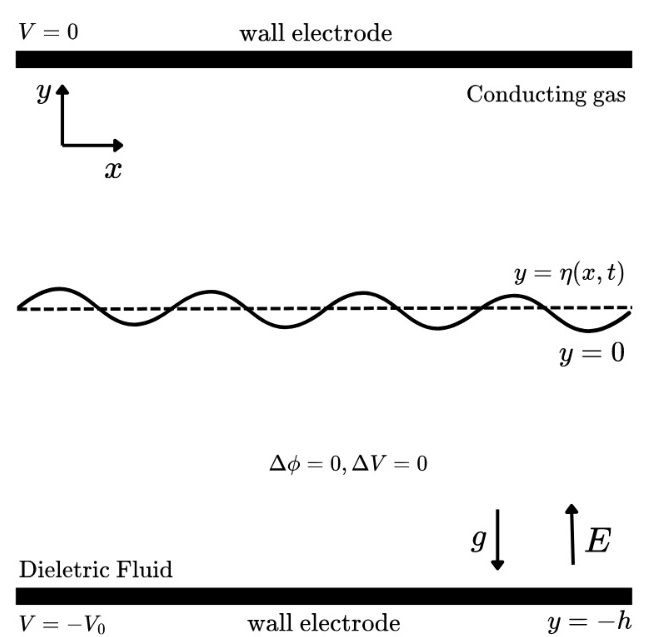}
	\caption{Sketch illustrating the physical context of the problem.}
	\label{fig_inicial}
\end{figure}

Let $\Vec{U} = \nabla \phi$ represent the irrotational velocity field of the fluid motion, where $\phi$ is a velocity potential. We denote the fluid surface by $\eta(x,t)$, which is assumed to be a periodic traveling wave of spatial period  $\lambda$  and speed $c$. {\color{black} A schematic is depicted in Figure \ref{fig_inicial}}.  By introducing the change of variables $X = x - ct$ and $Y = y$, we can express the fluid surface as $\eta(x,t) = \eta(X)$. Then, following the approach in \cite{FlamarionGaoRibeiroDoak:2022}, we can express the Euler governing equations in dimensionless form by selecting $h$, $\sqrt{h/g}$ and $V_0$ as the reference length, speed and voltage potential, in terms of the electric field and velocity potentials as follows

	\begin{align}
		\Delta \phi &= 0 \quad \text{in} \quad -1 < Y < \eta(X), \label{eq:1}\\
		\Delta V &= 0 \quad \text{in} \quad -1 < Y < \eta(X), \\
		-c \eta_X +\phi_X \eta_X &= \phi_Y  \quad \text{for} \quad Y = \eta(X), \label{eq:Kinematic}\\
		\phi_Y &= 0 \quad \text{for} \quad Y = -1, \\
		V &= 0 \quad \text{for} \quad Y = \eta(X), \\
		V &= -1 \quad \text{for} \quad Y = -1.
	\end{align}
	In addition to governing equations, we have the dynamic boundary condition:
\begin{equation}
	-c\phi_X + \frac{1}{2}(\phi_X^2 + \phi_Y^2) + \eta - \sigma \frac{\eta_{XX}}{(1+\eta_X^2)^{\frac{3}{2}}} + M_e = B,
\label{eq_boundary}
\end{equation}
where $B$ is the Bernoulli constant and $M_e$ represents the Maxwell stress tensor given by
\begin{equation}
	M_e = \frac{E_b}{2(1+\eta_X^2)}\Big[(1-\eta^2_X)(V_Y^2-V_X^2)-4\eta_X V_X V_Y\Big] = \frac{E_b}{2} |\nabla V |^2.
\end{equation}
The parameters $\sigma$ and $E_b$ represent the nondimensional Bond and Electric Bond numbers, respectively, defined as follows
\begin{equation}
	\sigma = \frac{T}{\rho g h^2}, \quad E_b = \frac{\epsilon_0 V_0^2}{\rho g h^3}.
\end{equation}

In the wave-moving frame, the trajectory \((X(t), Y(t))\) of a fluid particle in the dielectric fluid is governed by the dynamical system  
\begin{equation}
    \begin{dcases}
	\frac{dX}{dt} &= \phi_X(X,Y) - c,   \label{eq_sistphi1} \\
	\frac{dY}{dt} &= \phi_Y(X,Y). 
\end{dcases}
\end{equation}

In this frame, the trajectory coincides with the streamline of the flow.

\subsection{Stokes Drift} \label{subsec-drift}

In this section, we recall some properties and results on Stokes drift for water waves that are known in the literature.

{\color{black} For a solution \((X(t), Y(t))\) of the ODE system \eqref{eq_sistphi1} with the initial condition \((X_0, Y_0)\), where \( Y_0 \in [-1,\eta(X_0)] \), the time required for this solution to travel a single wavelength, i.e. reach $X=X_0-\lambda$,} is called the drift time and is denoted by \( \tau(Y_0) \). This represents the time that a particle takes to traverse one period in the moving frame.

 Constantin \cite{Constantin:2006}  proved the following  formula to calculate the drift time of a particle trajectory
\begin{equation}
\tau(Y_0) = \int_{-\lambda/2}^{\lambda/2} \frac{dx}{c - \phi_X (X,Y_0)}
\end{equation} 
Furthermore, Li and Yang  \cite{LiYang:2024} has shown that the drift time must satisfy the following inequality
\begin{equation}
\frac{\lambda}{c} < \tau(Y_0) \leq \int_{-\lambda/2}^{\lambda/2} \frac{1}{c - \phi_X(X, \eta(X))} dx,
\label{eq:desigualdade}
\end{equation}
where $\lambda/c$ is the wave period.

Following the established notation, we define the Stokes Drift as the distance between the points \( (x(0), y(0)) \) and \( (x(\tau(Y_0)), y(\tau(Y_0))) \), where \( x(t) = X(t) + ct \) and \( y(t) = Y(t) \).  As we can see at Li and Yang \cite{LiYang:2024}, a direct implication from \eqref{eq:desigualdade} is that the Stokes Drift is always positive.

\begin{figure}[h!]
	\centering
	\includegraphics[width=6cm]{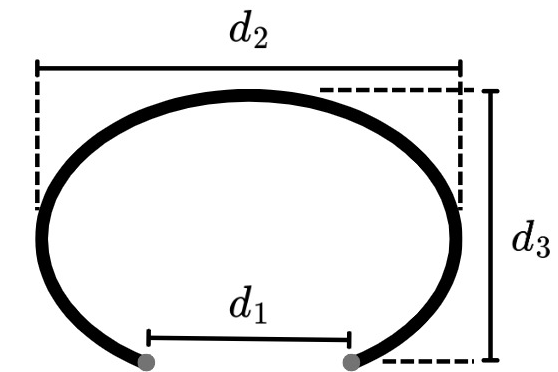}
	\caption{Geometric parameters indicating the aspect ratio of an trajectory.}
	\label{fig_d1d2d3}
\end{figure}

In our upcoming numerical experiments, we will investigate the properties of Stokes drift in the context of EHD flows. Additionally, we aim to analyze certain geometric characteristics of particle trajectories. To achieve this, we introduce the following three parameters:
\begin{enumerate}
	\item $d_1$ is the Stokes drift of the trajectory;
\item $d_2$ is the distance between $(x(t_1),y(t_1))$ and $(x(t_2),y(t_2))$ such that $t_1, t_2 \in [0,\tau(Y_0)]$, $\displaystyle x(t_1) = \min_{t \in [0,\tau(Y_0)]} x(t)$ and $\displaystyle x(t_2) = \max_{t \in [0,\tau(Y_0)]} x(t)$. It can be seen as the maximum horizontal distance.
\item $d_3$ is the distance between $(x(t_3),y(t_3))$ and $(x(t_4),y(t_4))$ such that $t_3, t_4 \in [0,\tau(Y_0)]$, $\displaystyle y(t_3) = \min_{t \in [0,\tau(Y_0)]} y(t)$ and $\displaystyle y(t_4) = \max_{t \in [0,\tau(Y_0)]} y(t)$. It can be seen as the maximum vertical distance.
\end{enumerate}
Figure \ref{fig_d1d2d3} presents a schematic representation of the parameters \( d_1 \), \( d_2 \), and \( d_3 \).

\section{Linear theory} \label{sec-linear}

In this section, we present the linear theory to derive an explicit formula for the velocity field of the dynamical system \eqref{eq_sistphi1}.

A trivial solution for the governing equations \eqref{eq:1}-\eqref{eq_boundary}  is 
\begin{equation*}
	\begin{cases}
		\eta_0(X) = 0, \\
		\phi_0(X, Y) = 0, \\
		V_0 (X, Y) = Y,
	\end{cases}
\end{equation*}
which is perturbed by a small disturbance measured by a parameter $\epsilon$ ($\ll 0$), namely
\begin{equation*}
	\begin{cases}
		\eta_0(X) = \varepsilon \hat{\eta}, \\
		\phi_0(X, Y) = \varepsilon \hat{\phi}, \\
		V_0 (X, Y) = Y + \varepsilon \hat{V}.
	\end{cases}
\end{equation*}
Here, $\varepsilon$ is a small parameter that measures the wave amplitude. Solving Laplace equations with boundary conditions as described in \eqref{eq:1}-\eqref{eq_boundary} yields
\begin{equation}
	\begin{cases}
		\hat{\eta}(X) = \mathfrak{R}\{Ae^{ikX}\}, \\
		\hat{\phi}(X, Y) = \mathfrak{R}\{Be^{ikX} \cosh (k (Y+1))  \}, \\
		\hat{V}(X, Y) = \mathfrak{R}\{Me^{ikX} \sinh (k (Y+1)) \}.
	\end{cases}
	\label{eq_chapeu}
\end{equation}
Here, $A$, $B$, and $M$ are unknown constants, and $k = 2 \pi / \lambda$ is the wavenumber. By linearizing the dynamic and kinematic boundary conditions, we obtain
\begin{equation*}
	\begin{cases}
		A = \varepsilon a, \\
		B = \frac{-iAc}{\sinh(k)}, \\
		M = \frac{-A}{\sinh(k)}, \\
	\end{cases}
\end{equation*}
the linear speed
\begin{equation} 
	c^2 = \tanh(k)\Big(\frac1k+\sigma k\Big) - E_b,
	\label{eq:c_linear}
\end{equation}
{\color{black} and  the linear velocity field 
\begin{align}
	\label{lin1}\hat{\phi}_X (X,Y) &= \frac{k A c \cos(kX) \cosh(k(Y+1))}{\sinh(k)}, \\
	\label{lin2}\hat{\phi}_Y (X,Y) &= \frac{k A c \sin(kX) \sinh(k(Y+1))}{\sinh(k)}.
\end{align}

From this point on, we only consider the positive value of $c$, i.e., the right moving waves.  Note that there exists a critical value \( E_b^* \)  such that waves are destabilized by the electric field when $E_b>E_b^*$  as the wave speed no longer admits a real solution. 
 In the particular case \( \lambda = 2\pi \) (i.e., \( k = 1 \)) and with zero surface tension (\( \sigma = 0 \)), we have \( E_b^* = \tanh(1) \). 
 To obtain a more accurate description of the limiting behavior of \( \tau \), which tends to infinity as \( E_b \to E_b^* \), a nonlinear theory is required.}




\section{The nonlinear problem: conformal mapping and numerical method} \label{sec-nonlinear}

To solve the dynamical system \eqref{eq_sistphi1}, we consider the velocity field solution of the full nonlinear Euler equations \eqref{eq:1}-\eqref{eq_boundary} by employing the conformal mapping formulation presented in \cite{FlamarionGaoRibeiroDoak:2022}, combined with the trajectory computation strategy outlined in \cite{Ribeiro:2017}. These approaches are based on the conformal mapping introduced by \cite{DyachenkoKuznetsovSpectorZakharov:1996}, which provided the foundation for the development of pseudospectral numerical methods for free-surface hydrodynamic problems in various contexts. 

The numerical methodology applied in our study consists of three main steps:

\begin{enumerate}
	\item Construct the conformal mapping suitable for the problem.
	\item Reformulate the Euler equations \eqref{eq:1}-\eqref{eq_boundary} in canonical coordinates. This reformulation allows us to determine a free-surface wave solution and the velocity field in the fluid domain for a given $E_b$.
	\item Solve the dynamical system \eqref{eq_sistphi1} in canonical coordinates, then map the corresponding trajectory back to the physical domain.
\end{enumerate}
For further details, we refer the reader to \cite{FlamarionGaoRibeiroDoak:2022} and \cite{Ribeiro:2017}. Here, we summarize only the key aspects of each step.

\begin{center}
{\it  1) Construct the conformal mapping suitable for the problem }

\end{center}

First, we construct a conformal mapping  
\[
Z(\xi,\zeta) = \tilde{X}(\xi,\zeta) + i \tilde{Y}(\xi,\zeta),
\]
which maps the strip of width \( D \) and length $L$,  
\[
\{(\xi,\zeta) \in \mathbb{R}^2 \mid -\frac{L}{2} < \xi < \frac{L}{2}, -D < \zeta < 0 \},
\]
onto the physical domain of a wavelength $\lambda$, 
\[
\{(X,Y) \in \mathbb{R}^2  \mid -\frac{\lambda}{2} < X < \frac{\lambda}{2}, -1 < Y < \eta(X) \}.
\]
This mapping flattens the free surface and satisfies the boundary conditions  
\[
\tilde{Y}(\xi, 0) = \eta(\tilde{X}(\xi, 0)), \quad \text{and} \quad \tilde{Y}(\xi, -D) = -1.
\]  

Using algebraic manipulations, the Fourier transform, and the assumption that \( Z \) is conformal, we obtain the following expressions:  

\begin{align}
	\tilde{X}(\xi,\zeta) &= -\mathcal{F}^{-1}_{k \neq 0} \left[ \frac{i \cosh(k(D + \zeta))}{\sinh(kD)} \mathcal{F}[{\mathbf{Y}}](k) \right] + \xi, \\  
	\tilde{Y}(\xi,\zeta) &= \mathcal{F}^{-1}_{k \neq 0} \left[ \frac{ \sinh(k(D + \zeta))}{\sinh(kD)}  \mathcal{F}[{\mathbf{Y}}](k) \right] +  \mathcal{F}[\mathbf{{Y}}](0) + \zeta, \label{sist_conformexy}  
\end{align}  
where \( \mathbf{Y}(\xi) = \eta(\tilde{X}(\xi, 0)) \) and \( \mathcal{F}[\cdot] \) denotes the Fourier transform:  

\begin{itemize}
	\item \( \mathcal{F}[f(\xi)] = \hat{f}(k_j) = \frac{1}{L} \int_{-L/2}^{L/2} f(\xi) e^{-ik_j\xi} \, d\xi \);  
	\item \( \mathcal{F}^{-1}[\hat{f}(k)] = f(\xi) = \sum_{j \in \mathbb{Z}} \hat{f}(k_j) e^{ik_j\xi} \);  
\end{itemize}  
where \( k_j = \frac{2\pi}{L} j \).  

\begin{center}
	{\it 2) Reformulating the Euler equations in canonical coordinates.}

\end{center}

We impose that the free surface has the same wavelength in both coordinate systems, i.e., \( L = \lambda \). This condition leads to the relation:  

\begin{equation}
	D = \frac{1}{L} \int_{-L/2}^{L/2} \mathbf{Y} (\xi) \, d\xi + 1.
	\label{eq_D}
\end{equation}  

Let \(\psi\) be the harmonic conjugate of \(\phi\), and define  
\[
\tilde{\phi}(\xi, \zeta) = \phi(\tilde{X} (\xi, \zeta), \tilde{Y}(\xi, \zeta)), \quad 
\tilde{\psi}(\xi, \zeta) = \psi(\tilde{X} (\xi, \zeta), \tilde{Y}(\xi, \zeta)).
\]  

Denoting \(\mathbf{X}(\xi)\) as the horizontal component of the conformal map evaluated at \(\zeta = 0\), the free surface in the canonical coordinate system corresponds to the curve \((\mathbf{X}(\xi), \mathbf{Y}(\xi))\). The kinematic \eqref{eq:Kinematic} and Bernoulli \eqref{eq_boundary} conditions reduce to:  

\begin{equation}
	\frac{-c^2}{2} + \frac{c^2}{2\mathbf{J}} + \mathbf{Y} + \sigma \frac{\mathbf{X}_\xi \mathbf{Y}_{\xi\xi} - \mathbf{Y}_\xi \mathbf{X}_{\xi\xi}}{\mathbf{J}^{3/2}} + \frac{E_b}{2D^2\mathbf{J}} - B = 0,
	\label{eq_G}
\end{equation}  
where \(\mathbf{J} = \mathbf{X}_\xi^2 + \mathbf{Y}_\xi^2\) is the Jacobian of the mapping evaluated at $\zeta = 0$.
We can rewrite Equation \eqref{eq_G} as
    \begin{equation}
   \frac12\Big(\frac{1}{J}-1\Big) \Big(c^2+\frac{E_b}{D^2}\Big) + \mathbf{Y} + \sigma \frac{\mathbf{X}_\xi \mathbf{Y}_{\xi\xi} - \mathbf{Y}_\xi \mathbf{X}_{\xi\xi}}{\mathbf{J}^{3/2}}  - B = 0.\label{eqn:eu}
      \end{equation}
In this way, for any electrified periodic wave solution with profile $\mathbf{Y}$ ($D$ the associated depth in the canonical plane) and speed $c$ in the presence of electric fields of strength $E_b$, the wave profile must be the same as a classic Stokes wave with speed $c_s$ where
      \begin{equation}
          c_s^2=c^2+\frac{E_b}{D^2}\,. \label{eqn:ceb}
      \end{equation}
Increasing the electric Bond number ($E_b$) would lead to a decrease in the value of the wave speed.

Equation \eqref{eq_G} involves four unknowns: $\mathbf{Y}$, $c$, $B$, and $D$.  To complete the system, we impose three additional conditions:
  \begin{enumerate}
	\item Fixed wave height:  
	\begin{equation}
		\mathbf{{Y}}(0) - \mathbf{{Y}}(L/2) = H.
		\label{eq_H}
	\end{equation}  
	\item Zero-mean wave profile in the physical space:  
	\begin{equation}
		\int_{-L/2}^{0} \mathbf{{Y}} \mathbf{{X}}_\xi \, d\xi = 0.
		\label{eq_media}
	\end{equation}  
	\item Depth condition \eqref{eq_D}.  
\end{enumerate}  

The equations \eqref{eq_G}, \eqref{eq_H}, \eqref{eq_media}, and \eqref{eq_D} are discretized spectrally, with derivatives computed via the Fast Fourier Transform (FFT) and integrals approximated using the trapezoidal rule. The resulting system is solved numerically using Newton's continuation method.  All calculations employ 1024 Fourier modes, with \( L = 2\pi \).

\begin{center}
{\it 3) Solve the dynamical system \eqref{eq_sistphi1} in canonical coordinates, then map the corresponding trajectory back to the physical domain. }  
\end{center}

	The trajectory \((X(t), Y(t))\), which is the solution of the ODE system \eqref{eq_sistphi1}, corresponds to the image of a trajectory \((\xi(t), \zeta(t))\) in the canonical domain under the conformal mapping. Specifically, we have  
		\begin{equation}
		(X(t), Y(t)) = (\tilde{X}(\xi(t), \zeta(t)), \tilde{Y}(\xi(t), \zeta(t))),
	\end{equation}  
		where \((\xi(t), \zeta(t))\) is determined by the system  
		\begin{equation}
        \begin{dcases}
            \frac{d \xi}{dt} &= \frac{1}{J} \left( \tilde{\phi}_\xi - c \tilde{Y}_\zeta \right), \label{eq:campo1} \\  
	\frac{d \zeta}{dt} &= \frac{1}{J} \left( \tilde{\phi}_\zeta + c \tilde{Y}_\xi \right).  
        \end{dcases}
	\end{equation}
		where \({J}(\xi,\zeta) = {\tilde{X}}_\xi^2 + {\tilde{Y}}_\xi^2\) is the Jacobian and  the velocity potential \(\tilde{\phi}(\xi, \zeta)\) is given by  
	
	\begin{equation}
		\tilde{\phi}(\xi,\zeta) = \mathcal{F}^{-1}_{k \neq 0} \left[ \frac{\cosh(k_j(D + \zeta))}{\cosh(k_jD)} \hat{\mathbf{\Phi}}(k) \right],
	\end{equation}
		with   	
	$	\mathbf{\Phi} = - \mathcal{C}[\mathbf{\Psi}]$,
	where the operator \(\mathcal{C}[\cdot]\) is defined as  
	$	\mathcal{C}[\cdot] := \mathcal{F}^{-1}[i \coth(k_j D ) \mathcal{F}[\cdot] ].$		Additionally, the stream function satisfies  
		$	\mathbf{\Psi} = c \mathbf{Y}.$

	Particle trajectories are computed numerically by integrating \eqref{eq:campo1} in the canonical domain using the fourth-order Runge-Kutta method. The resulting trajectories are then mapped onto the physical domain, yielding the trajectory  \((X(t), Y(t))\) in the moving frame. To analyze the particle paths in the laboratory frame, we apply the transformation  
	
	\begin{equation*}
		x(t) = X(t) + ct, \quad y(t) = Y(t).
	\end{equation*}

\section{Numerical results} \label{sec-results}

In this section, we present the results of numerical experiments for various electric Bond numbers ($E_b$).The objective is to understand the electric effect in the Stokes drift, so we ignore the surface tension by letting $\sigma = 0$. The results for non-zero capillarity are expected to be qualitatively similar. We also fix the depth at $1$, and set the wavelength ($\lambda$) to $2\pi$. We focus on this intermediate-depth regime and control the wave profile through the steepness parameter $\varepsilon = H / \lambda$. \textcolor{black}{We conduct a similar work for waves in shallow-water regime $\lambda  = 20\pi$ and deep-water regime $(\lambda = 0.2\pi)$ in Appendix \ref{appendix-a}.}

{\color{black} It is noted that, based on the linear wave speed from \eqref{eq:c_linear}, again with this set of parameters, the electric Bond number \( E_b \) must satisfy $E_b\leq E_b^*$ such that $c$ admits a real solution, in which $E_b^*=\tanh 1$ is the critical value as previously introduced.}

\subsection{Particle trajectories and Stokes drift}

\begin{figure}[h!]
    \centering
    \includegraphics[width=8cm]{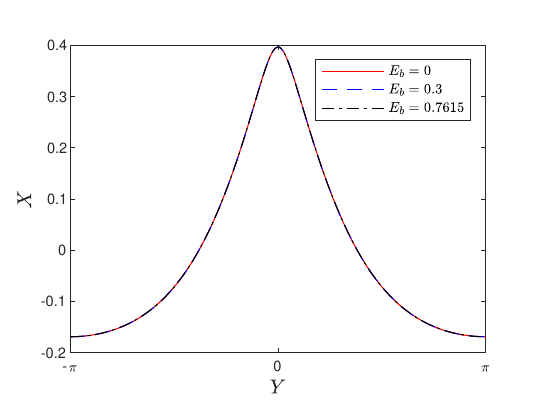}
    \caption{Wave profile for different values of $E_b$ for a fixed steepness $s=0.09$.}
    \label{fig_c1_xxeta_vareb}
\end{figure}

In this subsection, we explore the geometric parameters of the particle path, drift time, and wave speed for various values of $E_b$. Unless stated otherwise, the free-surface wave is fixed with a steepness of $s = 0.09$ \textcolor{black}{that guarantees us a strongly nonlinear wave profile with height $H \approx 0.5655$.} {\color{black}Similar to a remark made by \cite{CD:2018} for the case of infinite depth,  two electrified solutions may share the same wave profile as long as 
\begin{equation}
    c_1^2+\frac{E_{b1}}{D_1^2}=  c_2^2+\frac{E_{b2}}{D_2^2}\,,
\end{equation}
 due to equation \eqref{eqn:eu} and \eqref{eqn:ceb}. The value of the depth in the canonical domain does not vary provided the surface displacement is unchanged, and therefore, the main response to a change in $E_b$ may be reflected in the wave speed. This is confirmed by our numerical computations for three different values of $E_b$ as shown in Figure \ref{fig_c1_xxeta_vareb}, where the wave profiles for a fixed steepness look identical and the associated wave speed decreases as $E_b$ increases.} 
{\color{black} To further support this observation, we evaluated the pairwise distances \( \| \mathbf{Y}_i - \mathbf{Y}_j \|_2 \) between the solutions shown in Figure \ref{fig_c1_xxeta_vareb}.  The computed distances are of the order of \( 10^{-15} \), which is comparable  to numerical error. The numerical evidence shows that these three wave profiles are identical, and the difference is only subject to numerical error. }

The dynamical system \eqref{eq_sistphi1} is solved with the initial condition beneath the trough at $(\pi,Y_0)$. 

\begin{figure}[h!]
	\centering
	\includegraphics[width=6cm]{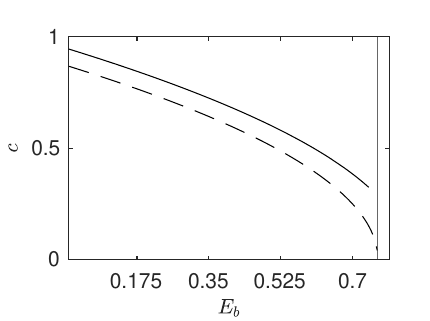}
	\includegraphics[width=6cm]{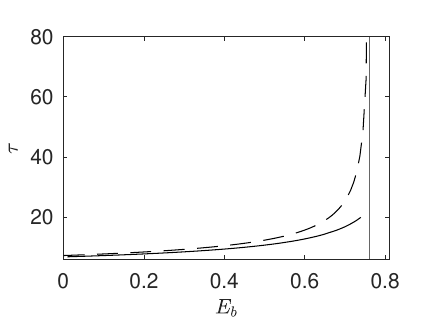}
	\caption{ {\color{black} Speed (\(c\)) and drift time (\(\tau\)) as functions of the electric Bond number (\(E_b\)). Results are shown for the linear theory (dashed lines) and the nonlinear theory (solid lines). Vertical lines indicate the critical value \(E_b = E_b^*\)}.
}
	\label{fig_speed_drifttime}
\end{figure}

\begin{figure}[h!]
	\centering
		\includegraphics[width=8cm]{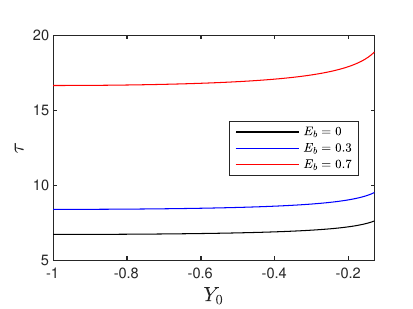}
	\caption{  Drift time  as a function of $Y_0$ for various values of $E_b$.}
	\label{fig_speed_drifttime_y0}
\end{figure}

The left graph in Figure \ref{fig_speed_drifttime} shows the wave speed (\(c\)), while the right graph shows the drift time (\(\tau\))  for various values of \(E_b\).  The particle's initial depth is $Y_0 \approx -0.5627$, which corresponds to a depth of  $\zeta = -D/2$ in the canonical domain.  {\color{black} {\color{black} For the linear solution, \( Y_0 = -0.5 \) (corresponding to \( (\xi_0, \zeta_0) = (\pi, -D/2) \))  and the trajectories are computed explicitly using linear theory}. } Note that, while the variation of $E_b$ does not affect the wave profile, the particle initial depth $Y_0$ remains identical regardless of $E_b$ value.  We observe that increasing \(E_b\) leads to a decrease in wave speed and an increase in drift time.  Additionally, Figure \ref{fig_speed_drifttime_y0}  shows that the drift time is larger near the free surface and decreases as we move from the trough toward the bottom, a result previously established for  for $E_b = 0$ \cite{LiYang:2024} and still valid for $E_b \neq 0$.

\begin{figure}[h!]
	\centering
	\includegraphics[width=6cm]{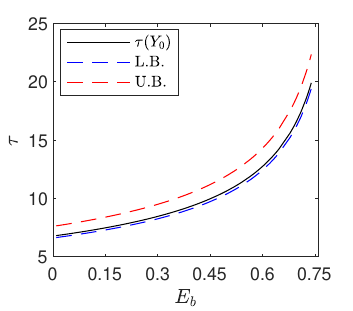}
	\includegraphics[width=6cm]{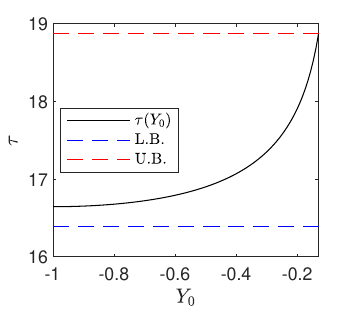}
	\caption{ Upper bound (UB) and lower bound (LB) of inequality \eqref{eq:desigualdade}, originally established for water waves, remain valid for EHD flows.    Left: $Y_0$ is fixed at $-0.5627$.  Right: $E_b$ is fixed at $0.7$.		 }
	\label{fig:LB_UB}
\end{figure}

We recall the inequality \eqref{eq:desigualdade}, which holds for water wave flow when $E_b = 0$. This inequality establishes lower and upper bounds for the drift time of a particle initially located at depth $Y_0$. As shown previously, when an electric field is introduced, the drift time also depends on the field intensity $E_b$. A natural question then arises: does inequality \eqref{eq:desigualdade} still hold when $E_b \neq 0$? 
The results are presented in Figure \ref{fig:LB_UB} where $Y_0$ is fixed at  $-0.5627$. In the left panel of that figure, the drift time $\tau(Y_0)$ is sketched for various values of $E_b$. We observe that $\tau(Y_0)$ remains bounded as predicted by inequality \eqref{eq:desigualdade}. The right panel of Figure \ref{fig:LB_UB} shows the drift time for different values of $Y_0$ in a flow with a fixed $E_b = 0.7$, confirming that inequality \eqref{eq:desigualdade} still holds.

\begin{figure}[h!]
	\centering
	\includegraphics[width=6cm]{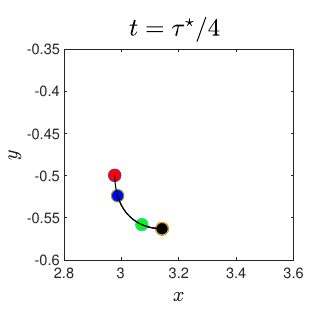}
	\includegraphics[width=6cm]{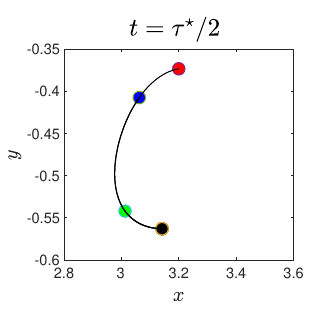}
	\includegraphics[width=6cm]{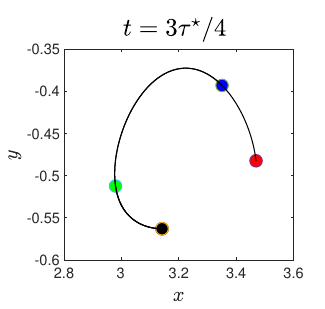}
	\includegraphics[width=6cm]{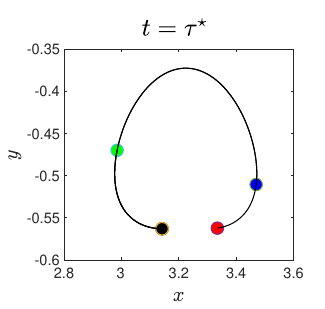}
	\caption{Snapshots of the trajectory of a particle starting from the same initial position, $(-\pi, -0.5)$, for the cases where $E_b = 0$ (red), $0.3$ (blue), and $0.7615$ (green). $\tau^\star$ represents the drift time of the particle for $E_b = 0$ and black dot the initial position of the particle. }
	\label{fig_snapshots}
\end{figure}

 Figure \ref{fig_snapshots} displays the trajectory of a particle starting from the same initial position,   $(\pi, -0.5627)$, for the cases where $E_b = 0$, $0.3$, and $0.7615$. We observe that higher values of $E_b$ result in slower particle motion.

\begin{figure}[h!]
	\centering
	\includegraphics[width=4cm]{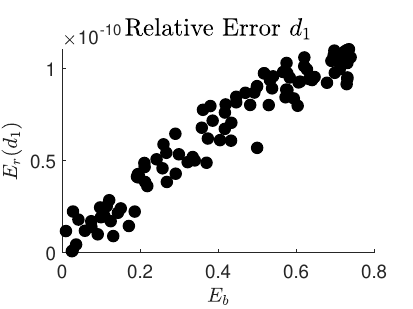}
	\includegraphics[width=4cm]{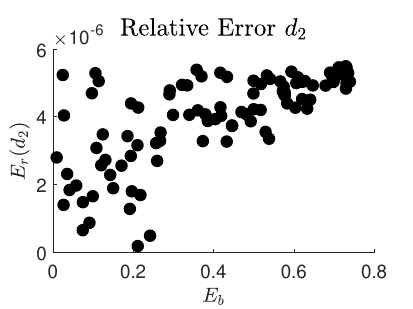}
	\includegraphics[width=4cm]{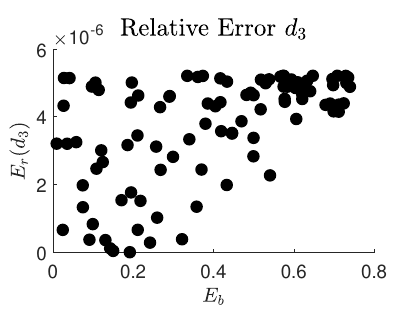}
	\caption{Relative errors of $d_1$, $d_2$ and $d_3$ versus $E_b$.}
	\label{fig_errorelativo}
\end{figure}

To analyze the geometric parameters \((d_1, d_2, d_3)\) of the particle trajectories, we first compute their values for the case \(E_b = 0\), denoted as \((d_1^\star, d_2^\star, d_3^\star)\), and use them as a reference to compare the results for other values of \(E_b\). The relative error is defined as  
\[
E_r(d_i) = \frac{|d_i - d_i^\star|}{|d_i^\star|}, \quad i = 1, 2, 3.
\]
As shown in Figure \ref{fig_errorelativo}, the relative errors in the geometric parameters \(d_1\), \(d_2\), and \(d_3\) remain small, which suggests that the trajectory shape is largely preserved as \(E_b\) increases, although the particle motion slows down.

\subsection{On the kinetic energy}  \label{sec-kinetic}

The total kinetic energy of a fluid particle initially located at \((X_0, Y_0)\) over a drift time in the moving frame is given by  
\begin{equation}
	E(X_0, Y_0) = \int_{0}^{\tau(Y_0)} \left[\left(\frac{dX}{dt}\right)^2 + \left(\frac{dY}{dt}\right)^2 \right] dt.
\end{equation}
By changing to the laboratory frame, where \(X = x - ct\) and \(Y = y\), the total kinetic energy over a drift time period in this frame is  
\begin{equation}
	\mathcal{E}(X_0, Y_0) = \int_{0}^{\tau(Y_0)} \left[\left(\frac{dX}{dt} + c\right)^2 + \left(\frac{dY}{dt}\right)^2 \right] dt.
\end{equation}

\begin{figure}[h!]
	\centering
	\includegraphics[width=6cm]{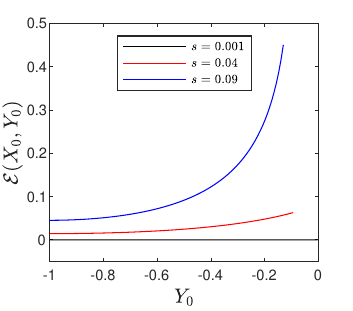}
	\includegraphics[width=6cm]{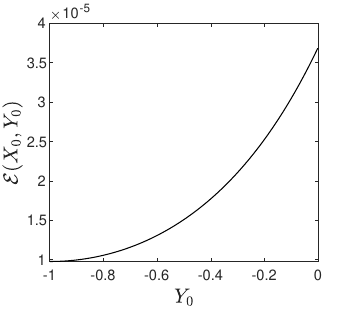}
	\includegraphics[width=6cm]{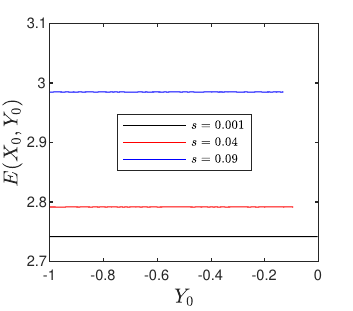}
	\caption{Top-left: Kinetic energy in the laboratory frame for various wave steepness values as a function of the particle's initial depth ($Y_0$). Top-right: a zoomed-in graph for $s = 0.001$. Bottom: Kinetic energy in the moving frame.}
	\label{fig_KE_3caso}
\end{figure}

For water wave flows, it is known that \cite{Olivia:2022, Olivia:2023, LiYang:2024}:  
\begin{enumerate}  
	\item \( E(X_0, Y_0) \) is a constant given by \( c \lambda / 2 \).  
	\item \( \mathcal{E}(X_0, Y_0) \) is a convex, non-decreasing function that depends solely on \( Y_0 \).  
\end{enumerate}

In the absence of an electric field, decreasing the wave steepness results in smaller kinetic energy as shown in Figure \ref{fig_KE_3caso}. For \(s = 0.001\), \(\mathcal{E}\) is significantly lower than in the other cases, warranting a zoomed-in figure for clarity.

\begin{figure}[h!]
	\centering
	\includegraphics{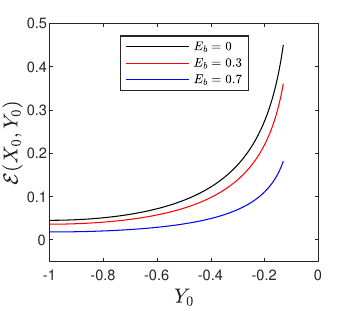}
	\includegraphics{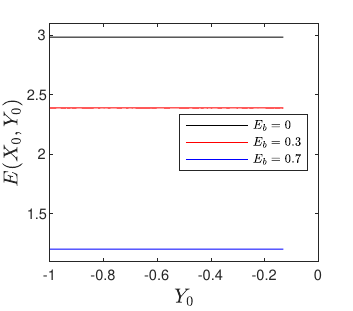}
	\caption{Kinetic energy in the laboratory frame (left) and the moving frame (right) as a function of the particle's initial depth ($Y_0$) for different values of $E_b$. }
	\label{fig_KE_2caso}
	\end{figure}

In addition, we want to numerically confirm the physical intuition that a decrease in steepness $s$ leads to lower kinetic energy in the context of EHD flows. To this end, we compute the total kinetic energy of 100 simulations, considering particles initially located at \( X_0 = \pi \) with \( Y_0 \) varying within the interval \( [-1, \eta(X_0)] \). Figure \ref{fig_KE_2caso} illustrates that varying \( E_b \) (\( E_b = 0, 0.3, 0.7 \)) preserves the known kinetic energy properties of water wave flows. Increasing \( E_b \) leads to a decrease in kinetic energy in both frames.

\section{Conclusion} \label{sec-conc}

In this work, we investigated the influence of the Electric Bound Number (\(E_b\)) on the Stokes Drift of a particle beneath a Stokes wave. Our numerical experiments showed that increasing \(E_b\) reduces wave speed and increases drift time. Additionally, the electric field slows down the particle trajectory while preserving its shape.  Besides, numerical results show that: (i) the inequality \eqref{eq:desigualdade}, and (ii) the property that drift time is larger near the free surface and decreases with increasing depth, both originally established for water waves, remain valid for EHD flows.

We also examined the kinetic energy properties of particle trajectories under periodic waves within the EHD flow framework. Our results indicate that kinetic energy is a non-decreasing convex function in the laboratory frame and remains constant in the moving frame, as its behaviour in water wave flows without an electric field.  
These findings contribute to a deeper understanding of particle transport mechanisms in EHD flows and provide insights into theoretical studies on the subject.

\section*{Acknowledgements}
The author L. P. P. is grateful for the financial support provided by CAPES Foundation (Coordination for the Improvement of Higher Education Personnel) during part of the development of this work. The work of M.V.F. was supported in part by the Direcci{\' o}n de Fomento de la Investigaci{\' o}n at the PUCP through grant DFI-2025-PI1277 and by National Council Scientific and Technological Development (CNPq) under Chamada CNPq/MCTI/No 10/2023-Universal. T.G. gratefully acknowledges the support of the London Mathematical Society with reference 42432 for sponsoring the visit of R.R. Jr. to the University of Essex, which made this work possible. T.G. would also like to thank the QJMAM Fund for Applied Mathematics for supporting the collaborative visit of  M. V. F. to the University of Essex.

\section*{Declarations}

\subsection*{Conflict of interest}
The authors state that there is no conflict of interest. 
\subsection*{Data availability}

Data sharing is not applicable to this article as all parameters used in the numerical experiments are informed in this paper.

\appendix 
\section{Shallow and deep-Water regime}\label{appendix-a}

\textcolor{black}{In this appendix, we aim to summarize the results obtained for waves in the shallow-water and deep-water regimes. We consider $\lambda_S = 10\pi$ and $\lambda_D = 0.5$ as wave length for shallow-water and deep-water regime, respectively. The same numerical methods previously applied to intermediate-water wave ($\lambda = 2\pi)$ regimes were used, and we investigated the variation of the drift time $\tau$ with $E_b$ variation, as well the geometric parameters of the orbit and the total kinetic energy of the particles.}

\begin{figure}[h!]
    \centering
    \includegraphics{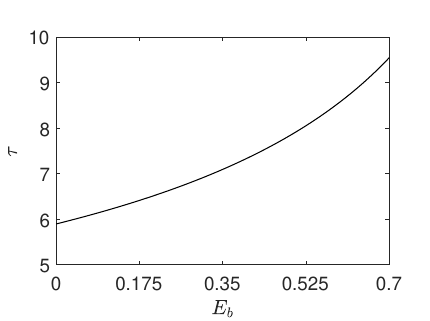}
    \includegraphics{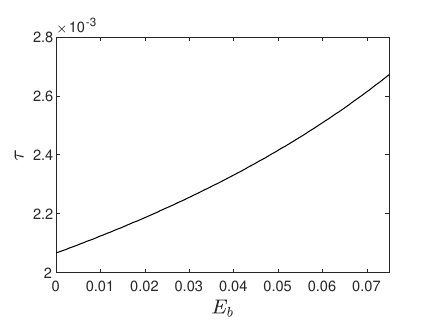}
    \caption{Drift time in function of $E_b$. Left: Shallow-water regime. Right: Deep-water regime.}
    \label{fig_shallow_deep_tdrift}
\end{figure}

\textcolor{black}{Our numerical experiments show analogous results for waves in intermediate-water regime. The drift time $\tau$ as function of the eletric bond number $E_b$ is crescent, as we can see in Figure \ref{fig_shallow_deep_tdrift}. Table \ref{tab_erros} shows the order of the relative error for the geometric parameters $d_1$, $d_2$ and $d_3$ in both regimes. The order of the relative error for the intermediate water wave regime is additionally presented in this table as a scale for comparison.}

\begin{table}[h!]
\centering
\begin{tabular}{|l|l|l|l|}
\hline
                   & $E_r(d_1)$ & $E_r(d_2)$  & $E_r(d_3)$ \\ \hline
Shallow-water      & $10^{-6}$  & $10^{-10}$  & $10^{-8}$  \\ \hline
Intermediate-water & $10^{-10}$ & $10^{-6}$   & $10^{-6}$  \\ \hline
Deep-water         & $10^{-6}$  & $10^{-10}$ & $10^{-10}$ \\ \hline
\end{tabular}
\caption{Order of the relative errors in three different regimes.}
\label{tab_erros}
\end{table}

\begin{figure}[h!]
    \centering
    \includegraphics{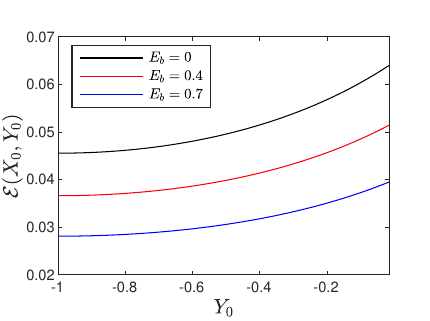}
    \includegraphics{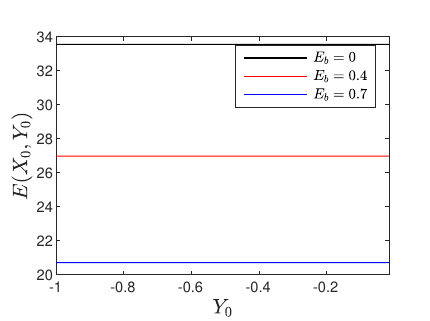}
        \includegraphics{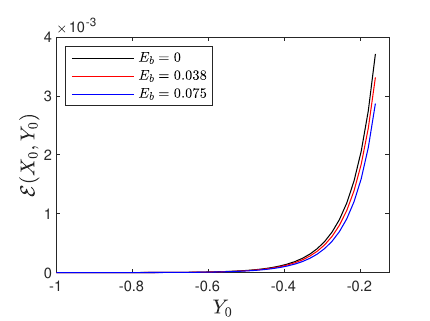}
    \includegraphics{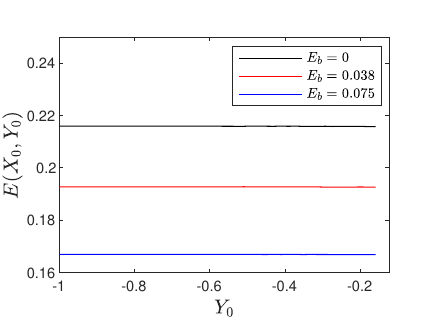}
    \caption{Total kinetic energy of particles as a function of the particle's initial depth ($Y_0$) for different values of $E_b$. Up: Shallow-water regime. Down: Deep-water regime  Left: Laboratory frame. Right: Moving frame. }
    \label{fig_deep_shallow_cinetica}
\end{figure}

\textcolor{black}{For the total kinetic energy of the particles, our results are presented in Figure \ref{fig_deep_shallow_cinetica} for shallow and deep-water regimes. As we can see in these figure, in laboratorial frame $\mathcal{E}(X_0,Y_0)$ is a convex and non-decreasing function and in moving frame $E(X_0,Y_0)$ is a constant function equal to $c\lambda/2$. Furthermore, the total kinetic energy of the particles in both regimes and frames of reference is lower for stronger electric fields.}

\begin{thebibliography}{999}

	
	\bibitem{Kalisch:2013}
	{\sc Alfatih A., Kalisch, H.,} Reconstruction of the pressure in long-wave models with constant vorticity. \textit{Eur. J. Mech. B Fluids} \textbf{37} (2013) 187-194.
	
	
	\bibitem{AbrashkinPelinovsky:2018}
	{\sc Abrashkin, A. A, Pelinovsky, E. N.,}  On the relation between Stokes drift and the Gerstner wave. \textit{Physics-Uspekhi} \textbf{61} (2018) 307-312.
	
	\bibitem{Airy:1841}
	{\sc Airy, G. B.,}  Tides and Wave. {\it Encyclopaedia Metropolitana} \textbf{5} (1842). 
	
	
			\bibitem{Mostafa:2015}
	{\sc 	Bakhoday-Paskyabi, M.,} Particle motions beneath irrotational water waves. {\it cean Dynamics}, 65.8 (2015): 1063-1078.
	
	

	\bibitem{Kalisch:2012}
	{\sc Borluk, H., Kalisch, H.,}  Particle dynamics in the KdV approximation. \textit{Wave Motion} \textbf{49}, (2012) 691-709.
	
	

	
	\bibitem{KalischCarterCurtis:2019}
	{\sc Carter, J. D., Curtis, C. W, Kalisch H.,}  Particle Trajectories in Nonlinear Schr{\"o}dinger Models. {\it Water Waves} \textbf{2} (2019) 31-57.
	
	\bibitem{Constantin:2006}
	{\sc Constantin, A.,}  The trajectories of particles in Stokes waves. {\it Invent. Math.} \textbf{166} (2006) 523-535.
	
	\bibitem{ConstantinVillari:2008}
	{\sc Constantin, A., Villari, G.,} Particle trajectories in linear water waves. {\it J. Math. Fluid Mech.} \textbf{10} (2008) 1336-1344.
	
	\bibitem{ConstantinStrauss:2010}
	{\sc Constantin, A., Strauss, W.,} Pressure beneath a Stokes wave. {\it Comm. Pure Appl. Math.} \textbf{63} (2008) 533-557.
	
	
	\bibitem{Olivia:2022}  
	{\sc Constantin, O.,} A complex-analytic approach to kinetic energy properties of irrotational traveling water waves. \textit{Mathematische Zeitschrift} \textbf{301} (2022) 4201-4215.
	
	
	\bibitem{Olivia:2023}  
	{\sc Constantin, O.,} A complex-analytic approach to streamline properties of deep-water Stokes waves. \textit{Arkiv f\"{o}r Matematik} \textbf{61} (2023) 81-97.
	
	
	
	
	
	
	
	
	
	\bibitem{Craik:2004}
	{\sc Craik, A.,} The origins of water wave theory. {\it Annu. Rev. Fluid Mech.} \textbf{36} (2004) 1-28.
	
	\bibitem{DyachenkoKuznetsovSpectorZakharov:1996}
	{\sc Dyachenko, A.I., Kuznetsov, E.A., Spector, M., Zakharov, V.E.,} Analytical description of the free surface dynamics of an ideal fluid
	(canonical formalism and conformal mapping). \textit{Phys. Lett. A} \textbf{221} (1996) 73-79. 
	
	\bibitem{FlamarionRibeiro:2021}
	{\sc Flamarion, M. V., Ribeiro-Jr, R.,} An iterative method to compute conformal mappings and their inverses in the context of water waves over topographies. \textit{Int. J. for Num. Methods in Fluids} \textbf{93} (2021) 3304-3311. 
	
	\bibitem{FlamarionGaoRibeiroDoak:2022}
	{\sc Flamarion, M. V., Gao, T., Ribeiro-Jr, R., Doak, A.,} Flow structure beneath periodic waves with constant vorticity under normal electric fields. \textit{Physics of Fluids} \textbf{34} (2022) 127119.
	
	\bibitem{Flamarion:2023}
	{\sc Flamarion, M. V.,} Complex flow structures beneath rotational depression solitary waves in gravity-capillary flows. \textit{Wave Motion} \textbf{117} (2023) 103108. 
	
	\bibitem{Flamarion:2023-2}
	{\sc Flamarion, M. V.,} Stagnation points beneath rotational solitary waves in gravity-capillary flows. \textit{Trends in Comp. and Apllied Math.} \textbf{24} (2023) 265-274. 
	
	\bibitem{FlamarionKochurinRibeiro:2023}
	{\sc Flamarion, M. V., Kochurin, E., Ribeiro-Jr, R.,}  Fully nonlinear evolution of free-surface waves with constant vorticity under horizontal electric fields. \textit{Mathematics} \textbf{11} (2023) 4467.
	
 	
	\bibitem{FlamarionGaoRibeiro:2023}
	{\sc Flamarion, M. V., Gao, T., Ribeiro-Jr, R.,}  An investigation of the flow structure beneath solitary waves with constant vorticity on a conducting fluid under normal electric fields. \textit{Physics of Fluids} \textbf{35} (2023) 037122.
	
	
	\bibitem{GhoshalMiner:2003}
	{\sc Ghoshal, U.,  Miner A. C.,} Cooling of high power density devices by electrically conducting fluids. \textit{U.S. Patent} (2003) 6,658,861.
	
	\bibitem{Green:1839}
	{\sc Green, G.,} On the motion of waves in a variable canal of small depth and width. \textit{Trans. of the Cambridge Phil. Soc.} \textbf{6} (1838) 457-462.
	
	\bibitem{KS:1997}
    {\sc S. F. Kistler and P. M. Schweizer}, Liquid Film Coating- Scientific Principles and their Technological Implications (Chapman and Hall, Springer, New York 1997).
    
	\bibitem{GriffingBankoffMiksisSchluter:2006}
	{\sc Griffing, E. M., Bankoff, S. G., Miksis, M. J., Schluter R. A.,}  Electrohydrodynamics of thin flowing films. \textit{J. Fluids Eng.} \textbf{128} (2006) 276-283.

    \bibitem{TD:1969}
     {\sc G. I. Taylor and M. D. Van Dyke}, Electrically driven jets, \emph{Proc. R. Soc.} A \textbf{313} (1969) 453-475.

\bibitem{CCY:2003}
{\sc X. Chen, J. Cheng and X. Yin}, Advances and applications of electrohydrodynamics, \emph{Chin. Sci. Bull.} \textbf{48} (2003) 1055-1063. 

\bibitem{P:2019}
{\sc D. T. Papageorgiou}, Film flows in the presence of electric fields, \emph{Ann. Rev. Fluid Mech.} \textbf{51} (2019) 155-187.

    \bibitem{TM:1965}
    {\sc G. Taylor and A. McEwan}, The stability of a horizontal fluid interface in a vertical electric field, \emph{J. Fluid Mech.} \textbf{22} (1) (1965) 1-15.
    
    \bibitem{MS:1968}
    {\sc J. R. Melcher and W.J. Schwarz}, Interfacial relaxation overstability in a tangential electric field, \emph{Phys. Fluids} \textbf{11} (12) (1968) 2604-2616.

\bibitem{D:2020}
        {\sc Doak, A., Gao, T., Vanden-Broeck, J. M., and Kandola, J. J. S.} (2020). Capillary-gravity waves on the interface of two dielectric fluid layers under normal electric fields. \emph{Quart. J. Mech. Appl. Math.}, \textbf{73}(3), 231-250.
        \bibitem{DC:1974}
    T. Perelman, A. Fridman and M. Eliashevich, Modified Korteweg-de Vries equation in electrohydrodynamics, \emph{Sov. Phys. JETP} \textbf{66} (1974) 1316-1323.
        \bibitem{DC:1988}
    C. Easwaran, Solitary waves on a conducting fluid layer, \emph{Phys. Fluids} \textbf{31} (1988) 3442-3443. 
    \bibitem{DC:2006}
    {\sc D. T. Papageorgiou and J.-M. Vanden-Broeck}, Numerical and analytical studies of nonlinear gravity–capillary waves in fluid layers under normal electric fields, \emph{IMA J. Appl. Math.} \textbf{72} (2006) 832-853.
    \bibitem{DC:2007}
{\sc H. Gleeson, P. Hammerton, D. Papageorgiou and J.-M. Vanden-Broeck}, A new application of the Korteweg-de Vries Benjamin-Ono equation in interfacial electrohydrodynamics, \emph{Phys. Fluids} \textbf{19} (2007) 031703. 
\bibitem{DC:2013}
{\sc P. Hammerton}, Existence of solitary travelling waves in interfacial electrohydrodynamics, \emph{Wave Motion} \textbf{50} (2013) 676-686. 
\bibitem{DC:2015} 
{\sc M. J. Hunt and J.-M. Vanden-Broeck}, A study of the effects of electric field on two-dimensional inviscid nonlinear free surface flows generated by moving disturbances, \emph{J. Eng. Math.} \textbf{92} (2015) 1-13.
\bibitem{DC:2017}
 {\sc Z. Wang}, Modelling nonlinear electrohydrodynamic surface waves over three-dimensional conducting fluids, \emph{Proc. R. Soc. A} \textbf{473} (2017) 20160817.
    \bibitem{DC:2022}
    {\sc Doak, A., Gao, T., and Vanden-Broeck, J.-M.} (2022). Global bifurcation of capillary-gravity dark solitary waves on the surface of a conducting fluid under normal electric fields. \emph{Quart. J. Mech. Appl. Math.}, \textbf{75}(3), 215-234.
\bibitem{CD:2018}
{\sc T. Gao, P. A. Milewski, D. T. Papageorgiou and J.-M. Vanden-Broeck}, Dynamics of fully nonlinear capillary-gravity solitary waves under normal electric fields, \emph{J. Eng. Math.} \textbf{108} (2018) 107-122. 
\bibitem{CD:2019}
{\sc T. Gao, A. Doak, J.-M. Vanden-Broeck and Z. Wang}, Capillary-gravity waves on a dielectric fluid of finite depth under normal electric field, \emph{Eur. J. Mech. B} \textbf{77} (2019) 98-107.
    
    \bibitem{LiYang:2024}
	{\sc Li, J., Yang, S.,} Kinetic energy properties of irrotational deep-water Stokes waves. \textit{arXiv:2406.00711v1. }
	
	\bibitem{Khorsand:2014}
	{\sc Khorsand, Z.,}  Particle trajectories in the Serre equations. {\it Appl. Math. Comput.}, \textbf{230} (2014) 35-42.
	
	
	\bibitem{Longuet-Higgins:1979}
	{\sc Longuet-Higgins, M. S.,} The trajectories of particles in steep, symmetric gravity. {\it J. Fluid Mech.} \textbf{94} (1979) 497-517.
	
	\bibitem{Longuet-Higgins:1986}
	{\sc Longuet-Higgins, M. S.,} Eulerian and Lagrangian aspects of surface waves. {\it J. Fluid Mech.}, \textbf{173} (1986) 683-707.
	
	
	

	
	\bibitem{NachbinRibeiro:2014}
	{\sc Nachbin, A., Ribeiro Jr, R.,} A boundary integral formulation for particle trajectories in stokes wave. {\it Disc. and Cont. Dynamical Systems} \textbf{34} (2014) 3135-3153.
	
	
	\bibitem{Ribeiro:2017}
	{\sc Ribeiro-Jr, R.,} {\sc Milewski, P. A.,} \& {\sc Nachbin, A.,} Flow structure beneath rotational water waves with stagnation points. {\it J. Fluid Mech.} \textbf{812} (2017) 792-814.
	
	
	\bibitem{Stokes:1847}
	{\sc Stokes, G. G.,} On the theory of oscillatory waves. {\it Trans. Cambridge Phil. Soc.} \textbf{8} (1880) 411-455.
	
	\bibitem{TelesDaSilvaPeregrine:1988}
	{\sc Teles da Silva, A. F., Peregrine, D. H.,}  Steep, steady surface waves on water of finite depth with constant vorticity. \textit{J. Fluid Mech} \textbf{195} (1988) 281-302.
	
	\bibitem{Ursell:1953}
	{\sc Ursell, F.,} Mass transport in gravity waves. \textit{Proc. Cambridge Phil. Soc.} \textbf{49} (1953) 145-150.
	
	\bibitem{VannesteYoung:2022}
	{\sc Vanneste, J., Young, W. R.,} Stokes drift and its discontents. {\it Phil. Trans. R. Soc. A} \textbf{380} (2022) 20210032.
	
	
	
		\bibitem{BremerBreivik:2018}
	{\sc van den Bremer, T. S., Breivik, {\o}.,}  Stokes Drift. \textit{Phil. Trans. of the R.S A: Math., Phys. and Eng. Sciences} \textbf{376} (2017) 20170104.
	
	\bibitem{Weber:2019}
	{\sc Weber, J. E. H.,} A Lagrangian study of internal Gerstner- and Stokes-type gravity waves. \textit{Wave Motion} \textbf{88} (2019) 257-264.
	
	
	
	

\end{thebibliography}
\end{document}